\documentclass[usenatbib,a4paper]{mn2e}
\usepackage{graphicx}
\usepackage{amssymb}
\usepackage{amsmath}
\usepackage{color}

\title{Are cold flows detectable with metal absorption lines?}
\author[Taysun Kimm et al.]{
\parbox[t]{\textwidth}{
Taysun Kimm$^{1}$\thanks{e-mail: taysun.kimm@astro.ox.ac.uk},
Adrianne Slyz$^{1}$, 
Julien Devriendt $^{1,2}$
and
Christophe Pichon$^{1,3}$
}
\vspace*{6pt} \\
$^1$ Department of Physics, Denys Wilkinson Building, Keble Road, Oxford, OX1 3RH, United Kingdom\\
$^2$ Centre de Recherche Astrophysique de Lyon, UMR 5574, 9 Avenue Charles Andre, F69561 Saint Genis Laval, France\\
$^3$ Institut d�astrophysique de Paris \& UPMC (UMR 7095), 98, bis boulevard Arago , 75 014, Paris, France
\\
}

\date{}                                           

\begin{document}
\maketitle

\newcommand{\msun}{\mbox{${\rm M}_{\odot}$}}
\newcommand{\mvir}{\mbox{$M_{\rm 200}$}}
\newcommand{\mstream}{\mbox{${M}_{\rm stream}$}}
\newcommand{\mn}{\mbox{\sc Horizon-MareNostrum}}
\newcommand{\nut}{\mbox{\sc Nut}}

\begin{abstract}

Cosmological simulations have shown that dark matter haloes are connected to each 
other by large-scale filamentary structures. Cold gas flowing within this ``cosmic web'' 
is believed to be an important source of fuel for star formation at high redshift. 
However, the presence of such filamentary gas has never been observationally confirmed 
despite the fact that its covering fraction within massive haloes at high redshift is 
predicted to be significant ($\sim$ 25\%), (Dekel et al. 2009). 
In this work, we investigate in detail whether such cold gas is detectable using 
low-ionisation metal absorption lines, such as CII $\lambda$1334 as this technique 
has a proven observational record for detecting gaseous structures.  Using a large 
statistical sample of galaxies from the Mare Nostrum N-body+AMR cosmological 
simulation, we find that the typical covering fraction of the dense, cold gas in 
10$^{12}$ M$_{\odot}$ haloes at z $\sim$ 2.5 is lower than expected ($\sim$ 5\%). 
In addition, the absorption signal by the interstellar medium of the galaxy itself 
turns out to be so deep and so broad in velocity space that it completely drowns 
that of the filamentary gas. A detectable signal might be obtained from a cold 
filament exactly aligned with the line of sight, but this configuration is so unlikely 
that it would require surveying an overwhelmingly large number of candidate 
galaxies to tease it out. Finally, the predicted metallicity of the cold gas in filaments 
is extremely low ($\leq$ 10$^{-3}$ Z$_\odot$). 
Should this result persist when higher resolution runs are performed, 
it would significantly increase the difficulty of detecting filamentary gas inflows 
using metal lines. However, even if we assume that filaments are enriched to Z$_\odot$, 
the absorption signal that we compute is still weak. We are therefore led to conclude 
that it is extremely difficult to observationally prove or disprove the presence of 
cold filaments as the favorite accretion mode of galaxies using low-ionisation metal 
absorption lines. The Ly$\alpha$ emission route looks more promising but due to the 
resonant nature of the line, radiative transfer simulations are required to fully 
characterize the observed signal. 

\end{abstract}

\voffset=-0.6in
\hoffset=0.2in

\section{Introduction}

How galaxies get their gas is a long-standing issue. 
For decades, the standard theoretical picture of galaxy formation has stipulated that  
all gas accreted into dark matter haloes is shock heated before it radiatively cools
and settles into a galactic disk \citep[][although \citealt{binney77} first suggested 
this need not be the case]{silk77,rees77}. This picture has recently been 
revisited both by analytic studies \citep{birnboim03,dekel06} and 
hydrodynamical simulations (both in 1D: \citealt{birnboim03} and in 3D within an 
explicit cosmological context, \citealt{keres05,keres09,ocvirk08}; hereafter OPT08). 
These studies have established that in haloes below a critical mass shocks are unstable 
and cannot propagate outwards, so that cold diffuse gas and/or cold filaments can 
penetrate deep into the halo without experiencing shock-heating.
In contrast, at the other end of the mass spectrum, very massive haloes easily sustain 
a virial shock that is stable against gas cooling so that diffuse/filamentary 
gas is shock heated to the virial temperature of the halo as it enters. 
Finally, at intermediate halo masses, either gravitationally shock-heated hot 
gas and/or a hot galactic wind coexists with cold inflowing filaments
(OPT08): some dense cold filaments are stable against the pressure force 
exerted by the hot gaseous material. In other words, the vast majority of galaxy-size host halos, 
especially at high redshift ($z>2$), are predicted to be threaded by cold gas filaments 
in a $\Lambda$CDM model of structure formation. Therefore, the question which naturally 
arises is whether or not the existence of these cold gas filaments can be observationally
confirmed, and by which technique.

Since the advent of high-resolution spectroscopy has enabled astronomers to study the 
kinematics of the intergalactic medium at high redshifts in a wealth of detail 
\citep[e.g.][]{pettini02,adelberger03,shapley03}, and given the fact that cold gas is 
thought to flow into massive haloes ($10^{12} \msun$) at $z=2.5$ along filaments with 
velocities of $\gtrsim 200\,{\rm km s^{-1}}$ \citep{dekel09}, it is sensible to think 
that the spectra of Lyman-break galaxies \citep{steidel96} might reveal these filaments 
as redshifted absorption features. Interestingly, a recent study reported that few 
Lyman-break galaxies (LBGs) show redshifted metal absorption lines, suggesting that 
inflowing gas in haloes with $4\times10^{11}<M_{\rm vir}<10^{12}M_{\odot}$ is rare 
\citep{steidel10}. Taken at face value, this appears to contradict the theoretical 
prediction that cold filaments are prominent in the vast majority of high-z halos. 

However, there exist a variety of reasons as to why these filaments should be very 
difficult to detect. The first of these is the covering fraction of the filaments. 
This is estimated in \citep{dekel09} to be around ($\sim 25\%$) for four massive 
halos with $M_{\rm vir}\sim10^{12}M_{\odot}$ in the 
\mn\ simulation, counting only relatively dense ($N_{{\rm H}}>10^{20}{\rm cm}^{-2}$) 
and cold ($T<10^{5}K$) gas within a radial distance $20<r<100$ kpc from the central 
galaxy hosted by these halos. Whereas $25\%$ is indeed a non-negligible covering 
fraction, its exact dependence on redshift and halo mass remains to be determined. 
 For instance, \citet{faucher10a} also measured the covering fraction
 for a Milky way-type progenitor LBG in a smoothed particle hydrodynamic simulation 
 but found a significantly smaller value ($\sim 2\%$). 
 Secondly, low-ionisation lines can be produced  not only by cold filamentary gas 
 but also by a galaxy's interstellar medium, so that when using a single galaxy to 
 probe the circumgalactic medium, distinguishing absorption produced by filaments
from that produced by the ISM is key to prove/disprove the presence of cold filaments. 
Thirdly, there is the geometry of the accretion: in order to produce a strong
absorption signal, a filament needs to be well aligned with the line
of sight to maximise its column density. Finally, there is the issue of metallicity: 
a metal-poor filament will be transparent to metal line observations.

The aim of this letter is to quantify the aforementioned effects to
better assess the detectability of the absorption signal 
produced by cold filaments. We show that the actual probability of detecting such flows 
with metal lines is much smaller than the high covering fraction
derived in \citet{dekel09} would suggest due to  
i) the low density and ii) low metallicity of the filaments 
{\em compared with the densities and metallicites of the interstellar
  medium of the host galaxy. }

\section{Simulations}

To investigate the statistical properties of cold filaments, we analyze 
the \mn\ simulation, carried out with the {\sc WMAP1} cosmology 
\citep{spergel03} using the adaptive mesh refinement code {\sc ramses} \citep{teyssier02}.
Details of the simulation can be found in OPT08, \citet{dekel09},
\citet{devriendt10}, so we just briefly describe the simulation setup
and modelling strategy here.

The simulation follows the evolution of a periodic cosmological volume of comoving 
side length $L=50\,{h^{-1}}$Mpc, and contains $1024^3$ dark matter particles 
with $m_{\rm dm}=1.4\times10^7 {\rm M}_{\odot}$. Dark matter halos are identified 
using the {\sc AdaptaHop} algorithm \citep{aubert04,tweed09}, resulting in 3419 halos 
with $M_{\rm vir} \ge 10^{11} {\rm M}_\odot$ at $z=3.8$ and 6456 haloes at $z=1.5$.
The spatial resolution of the simulation is kept fixed at around $1\,h^{-1}$kpc physical
over the entire redshift range.

Gas in the simulation can radiatively cool by atomic processes down to $10^4 K$ 
\citep{sutherland93}. A fraction of the cold and dense gas ($n_{\rm H}>0.1 {\rm cm}^{-3}$)  
turns into stars following the Schmidt-Kennicutt law \citep{kennicutt98},
and massive stars explode as type II supernovae, redistributing energy and metals 
into the interstellar/intergalactic medium. The Sedov Blast wave solution is adopted 
for supernova explosions \citep{dubois08}, and reionization is 
implemented by instantaneously turning on a uniform UV 
background at z=8.5 \citep{haardt96}.

 \section{Results}
 \subsection{Covering Fraction of Dense Gas}

\begin{figure} 
   \centering
   \includegraphics[width=8.8cm]{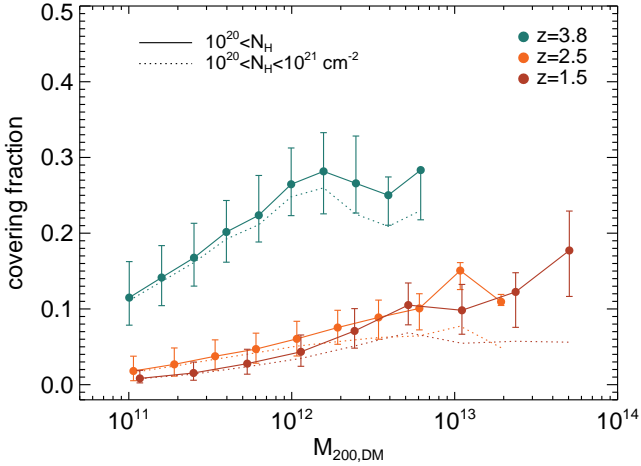} 
  \caption{The covering fraction of cold ($T<10^5 K$) and dense ($N_{\rm H} > 10^{20}$) gas within 
   $20 < r < 100$ kpc physical as a function of the virial mass of halos (\mvir). 
   Different colours indicate the covering fractions at different redshifts. Solid lines 
   include the contribution from the interstellar medium of satellite
   galaxies to the covering fraction. To exclude this latter contribution, we also plot the covering fraction 
   with upper density cut ($10^{20}<N_{\rm H}<10^{21}$ cm$^{-2}$, dotted lines). Error bars correspond 
   to the interquartile range ($25\le f \le 75\%$). For a given halo mass, halos at higher 
   redshift show larger covering fractions. This can be understood in
   terms of the rapid development of a virialized hot medium between $z=3.8$ and $z=2.5$. 
   }
   \label{fig:covfrac}
\end{figure}

OPT08 showed that the transition mass (\mstream)
separating cold from hot dominated accretion increases with increasing redshift.
According to these authors, by $z\sim2.5$, the cold streams feeding massive haloes 
($\mvir \gtrsim 3\times10^{11}\msun$) begin to disappear, but based on measurements 
of a handful of haloes in the \mn\ simulation at $z\sim2.5$, \citet{dekel09} 
find that some massive haloes with $\mvir\sim10^{12}\msun$  
still show high covering fractions ($\lesssim 25\%$) of dense ($N_{\rm H}>10^{20} {\rm cm}^{-2}$) 
and cold ($\rm T<10^5 K$)  inflowing gas within $20 < r < 100$ kpc. 
In order to obtain statistically significant results on the covering fraction 
in the \mn\ simulation, we analyse {\em all} the halos in the simulation and show 
the results in Fig. \ref{fig:covfrac}. Using the complete sample, we find that the 
average covering fraction of haloes with $\mvir\sim10^{12}\msun$ at $z=2.5$ is $\sim 5\%$, 
about a factor 5 less than the covering fraction reported in \citet{dekel09} 
for their sub-sample of $\mvir\sim10^{12}\msun$ halos at $z\sim2.5$.
We note that this lower value is more consistent with the recent findings of \citet{faucher10a}. 
However, whilst low covering fractions ($\sim 5\%$) are computed for most halos at redshifts
$z\lesssim 3$, those of higher redshift ($z\sim 3.8$) halos are much larger ($\sim 25\%$), 
as can be seen in Fig. \ref{fig:covfrac}. This strong redshift evolution in the covering 
fraction of cold filaments between $z\sim3.8$ and $z\sim2.5$ reflects the aforementioned 
rapid transition from cold to hot dominated accretion. We note that these results are 
consistent with the OPT08 value  of $\sim 10^{13}\msun$ for \mstream\ at $z=4$.

An interesting feature present in Fig. \ref{fig:covfrac} is that the 
covering fraction is higher in more massive haloes at a given redshift. 
This seems to contradict previous findings (OPT08) that it is the small haloes which are mainly 
fed by cold mode accretion. It should be noted, however, that the covering fraction 
shown in Fig. \ref{fig:covfrac} does not account for the accretion of cold, {\em diffuse} 
gas (cold gas with lower column densities) which is only present in halos with masses 
incapable of sustaining a virial shock at all. Indeed, these small haloes are usually 
located within (or around) filaments whose density is low, whereas the filaments around 
more massive haloes tend to be denser. Furthermore, in massive haloes,  satellite galaxies 
contribute more importantly to the covering fraction, as do extended, warped, galactic 
disks and dense gas bridges which result from tidal interactions between galaxies.
These latter effects partly explain the trend, but the primary driver of the covering 
fraction increase with halo mass is the density of the accreted gas, which is higher in more 
massive haloes.  Fig. \ref{fig:covfrac} (dotted lines) substantiates this claim by showing 
how the covering fractions drop when the very dense gas ($\rm N_H >
10^{21} cm^{-2}$) which belongs to the ISM of satellite galaxies is excluded from the measurement.

 \subsection{CII Absorption}
 
  \begin{figure} 
   \centering
   \includegraphics[width=8cm]{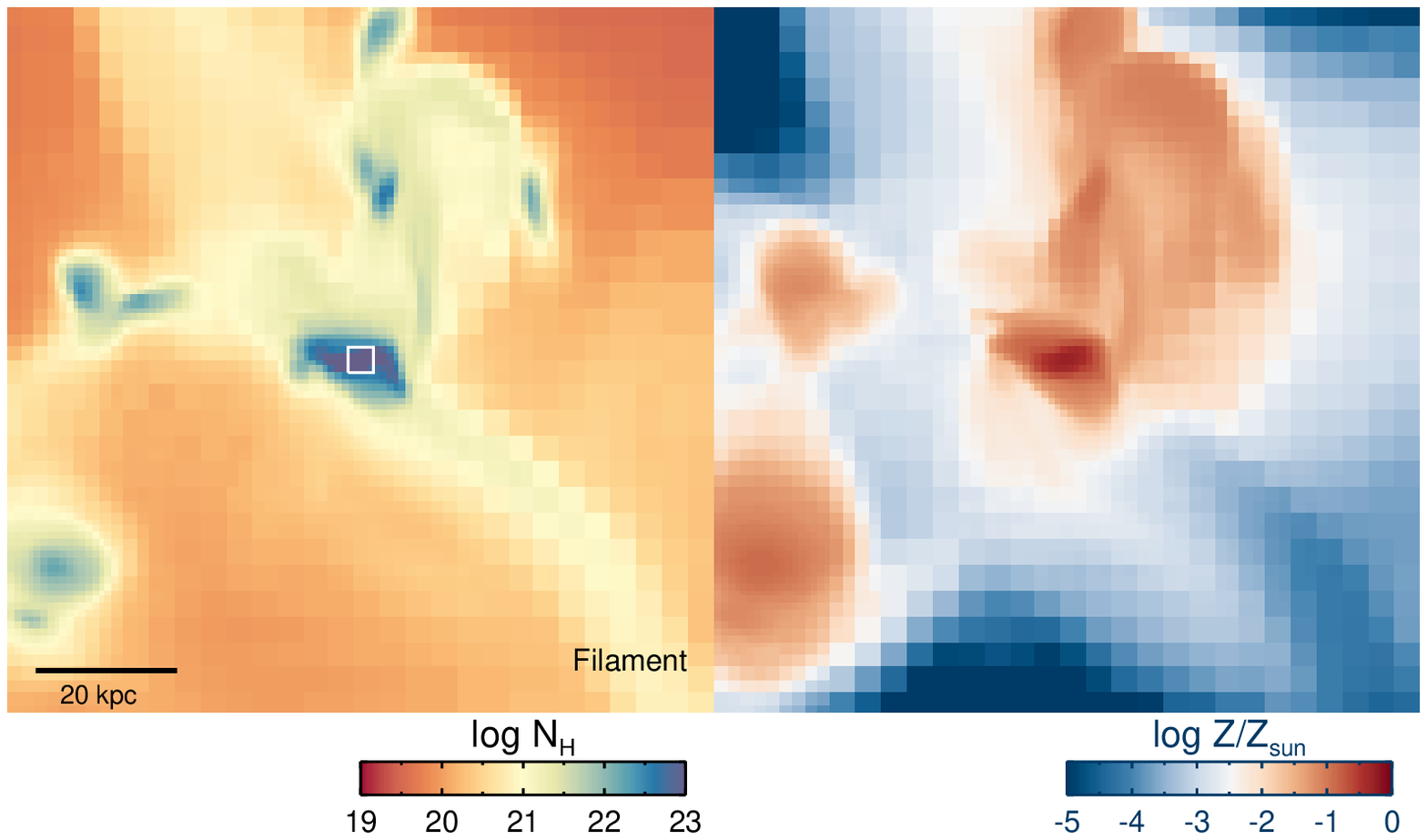} 
   \includegraphics[width=8.5cm]{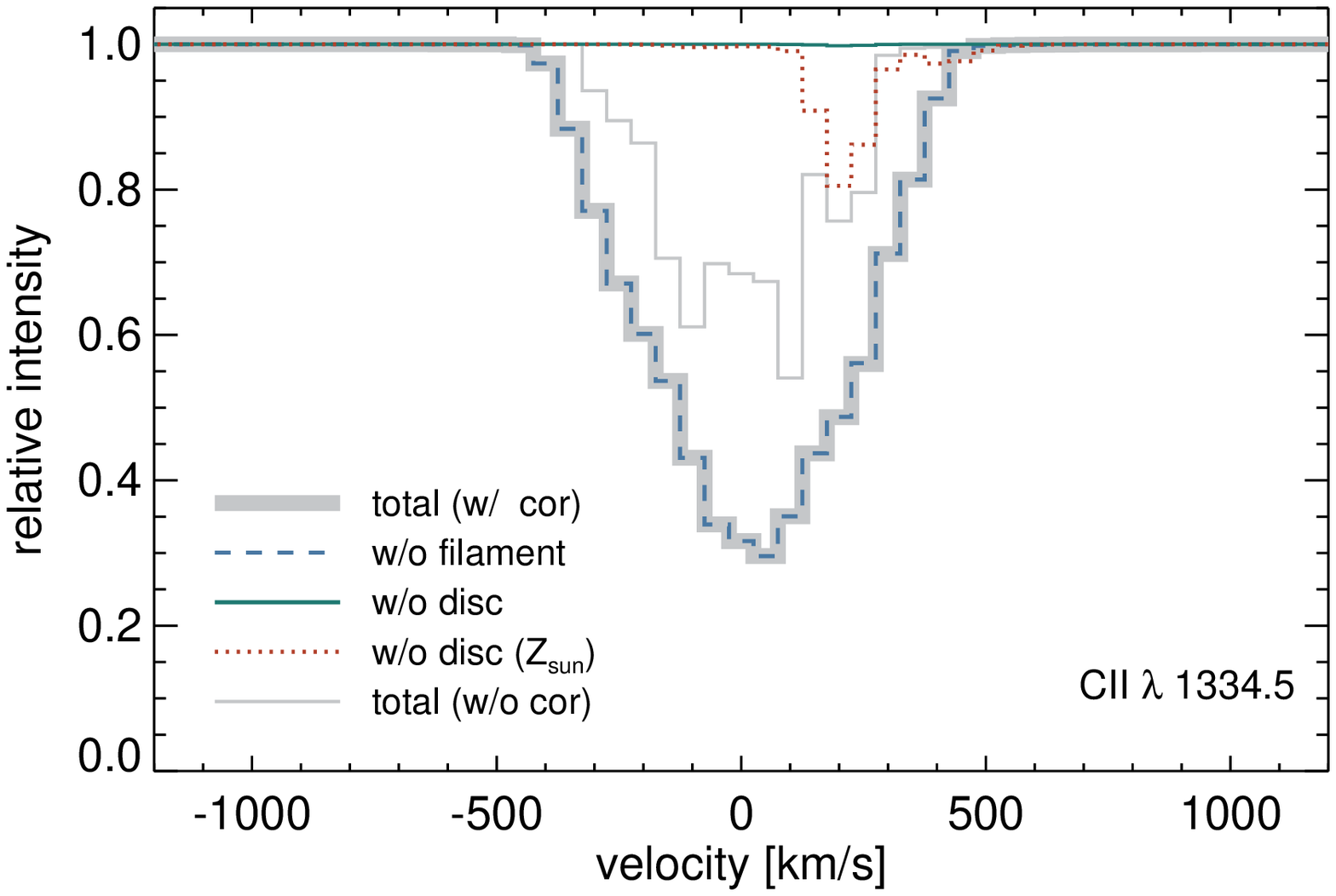} 

   \caption{An example of the contribution from filamentary gas to the CII $\lambda1334$ 
   absorption. Upper panels show the hydrogen column density and mass-weighted 
   metallicity distributions for a $10^{12}\msun$ halo at
   $z\sim3.8$. A cold filament is indicated on the 
   left panel, which is receding from  the observer and therefore
   should produce a redshifted absorption line. A white square denotes the central $\sim$10 kpc$^2$ 
   region over which the absorption spectrum is obtained. Bottom panel shows an absorption profile
   of this galaxy (thick grey solid line). We also compute the absorption without the central disc
   by neglecting the opacity of central cells ($r\leq0.1r_{200}$) (green solid line) and 
   absorption without the outer gas ($r>0.1r_{200}$) (blue dashed line). Also included is the 
   absorption produced when the outer gas ($r > 0.1 r_{200}$) is assumed to have solar metallicity in 
   the absence of the central disc (red dotted line). It can be seen that the contribution from the 
   filament to the low-ionisation metal line is negligible compared to
   that of the ISM of the galaxy in every case. An uncorrected spectrum is included to show the effect of the correction applied to the velocity distribution of gas (thin solid grey line).}
   \label{fig:example}
\end{figure}
 
In order to more carefully investigate the possibility of detecting cold filaments using 
metal absorption lines, we compute the strength of the ${\rm CII} ~\lambda 1334$ 
absorption. Our choice of line is dictated not only by the temperature of the filamentary gas which 
is not high enough to significantly produce more highly ionized metallic elements such as CIV,
but also because it is empirically known to yield the strongest absorption feature \citep{steidel10}
We do not attempt to model absorption by CII accurately which would require detailed radiative 
transfer, but instead derive an upper limit by making several extreme assumptions. Firstly, 
we assume that all the carbon present in filaments is eligible for the ${\rm C II}~ \lambda 1334$ 
transition. Secondly, we use the solar abundance ratio ($[C/Z]_{\odot}\simeq 0.178$, 
\citealt{asplund09}) to obtain the carbon column density for a given metallicity in the simulation. 
The optical depth of a grid cell is computed as $\tau  = \sigma_{\rm C II} n_{\rm C II} \Delta l$, 
where $n_{\rm CII}$ is the carbon number density, $\Delta l$ is the size of the grid cell, 
and $\sigma_{\rm C II}$ is the cross-section for the line transition, which is calculated as
$\sigma_{\rm C II}=(3\pi \sigma_{\rm T}/8)^{1/2} f \lambda_0\simeq 1.5 \times 10^{-18} {\rm cm}^2$.
Here $\sigma_{\rm T}$ is the Thomson cross-section, $\lambda_0$ is the rest-frame wavelength of the 
transition, and $f$ is the corresponding oscillator strength. We then correct the optical depth  
by assuming  that each grid cell has a Gaussian velocity distribution with a dispersion 
($\sigma_{\rm 1D}/2$), which is obtained from the line-of-sight velocity dispersion 
($\sigma_{\rm los}$) computed using the closest 27 neighbouring cells. 
We have tested the validity of the correction by using a high resolution in the \nut\ series 
(12 pc resolution, Devriendt  et al. {\sl in prep.}),  and found that this procedure yields a accurate 
approximation of the maximum absorption strength 
and FWHM of the absorption profile that one would derive by using a much higher number of resolution elements. 
Finally stars are assumed to dominate the UV emission and we use the \citet{maraston05} spectral 
energy distributions to derive the continuum flux around $\lambda\sim1334\AA$, which depends on 
their mass, age, and metallicity. The emission from each star particle in the galaxy is then used to 
estimate an observed flux, as attenuated by intervening gas present along the line of sight. Note that only the 
flux emitted by the central 10 kpc$^2$ of a galaxy is used to compute the absorption line so as to 
mimic the observational resolution of FWHM$\simeq 0.40''$ for galaxies at $z\sim2$ \citep{steidel10}.

Fig. \ref{fig:example} shows the HI column density map, projected metallicity distribution, 
and the corresponding absorption profile for a galaxy residing in a $\mvir \sim10^{12}\msun$ 
halo at $z\sim3.8$. A dense filament is clearly seen not only in the H column density map, 
but also in the metallicity map, with metallicity  ($\sim 10^{-4}-10^{-3} Z_{\odot}$).  
The filamentary gas is receding from the observer; hence if detectable it should produce a 
redshifted absorption line. However, it turns out that the absorption signal is dominated 
by the ISM of the galaxy lighting up the filament. When the absorption due to the ISM is 
arbitrarily removed by neglecting the opacity from the gas inside the central gas disc 
($r<0.1 r_{200}$),  the absorption feature vanishes (green line). Even when the gas outside 
$r > 0.1 r_{200}$ is assumed to have solar metallicity, the absorption
strength is still much smaller (red dotted line) than the absorption produced 
by the galaxy's ISM (grey and blue dashed lines). This strongly suggests that the 
primary reason why it is so difficult to detect the cold filament  is that the density of the 
filamentary gas is much lower than that of the galaxy's own ISM. As a result, and since 
the two absorption lines are not well enough separated in velocity space, 
the filament absorption signal is completely swamped by the high level of 
ISM absorption in the red wing of the line. 

\begin{figure} 
   \centering
   \includegraphics[width=8.5cm]{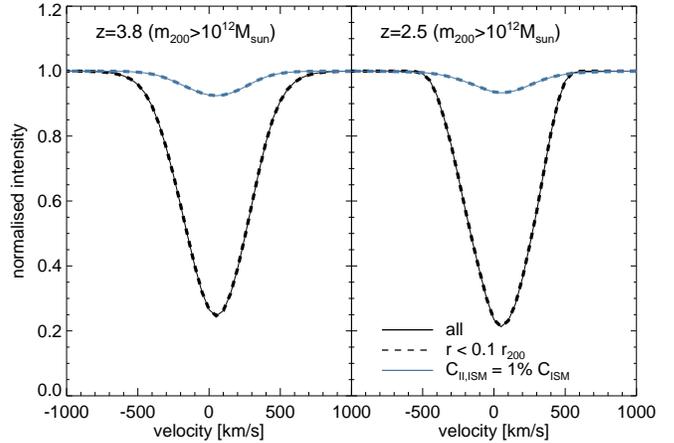} 

   \caption{Stacked absorption profiles of massive galaxies ($M_{200}\ge 10^{12}\msun$) at 
   $z=3.8$ (left) and $z=2.5$ (right). We take 6 projections ($+x, -x, +y, -y, +z, -z$) for 132 
   galaxies at $z=3.8$ and 386 galaxies at $z=1.5$. The stacked spectra are obtained by taking 
   the mean of the normalised intensity. We also show the absorption spectra expected 
   when only a fraction (1\%) of the ISM carbon is assumed to produce the absorption (blue lines, see the text). 
   The contribution from the outer regions (i.e. filament if any) is negligible in the 
   absorption profile. }
   \label{fig:stat}
\end{figure}

In order to see if we can bring out the filamentary signal by stacking absorption profiles, 
we analyse the absorption spectra of 132 and 386 massive galaxies ($M_{200}\ge 10^{12}\msun$) 
at $z=3.8$ and $z=2.5$ respectively along 6 projections ($+x, -x, +y, -y, +z, -z$) and 
find that the optical depth for the CII $\lambda 1334$ transition by the filaments is 
fundamentally small regardless of redshift. Fig. \ref{fig:stat} shows that the stacked
(mean) absorption strength is unaffected by the presence of filaments. 
To check whether the signal from filaments could potentially become noticeable if the 
ISM was less metal-enriched, we also examined the case  where only a tiny fraction (1\%) 
of the carbon in the ISM is eligible for the CII transition (blue lines in Fig. \ref{fig:stat}), 
but the difference between absorption profiles resulting from all the gas along the 
line of sight versus the case where we exclude the gas in the central region ($r<0.1 r_{200}$) is still minute. 
Therefore, we conclude that the presence of cold filaments is very difficult to confirm 
with low-ionisation metal absorption lines.

The optical depth of filaments may be under-estimated due to the finite 
resolution of the \mn\  simulation. For example, the density of the filamentary structure 
at $z=7$ in the \mn\  simulation ($0.005 \lesssim n_{\rm H} \lesssim 0.1$)
is more than an order of 
magnitude lower than that of the \nut\ simulation ($0.1 \lesssim n_{\rm H} \lesssim 1$) 
where the filaments are fully resolved (Powell et al. {\sl in prep.}). However, this increase does not suffice to 
produce a strong absorption signal. Moreover, we find that the effect of increasing 
resolution affects the ISM density more considerably (it becomes more than four orders 
of magnitude higher in the \nut\ than in the \mn\ simulation). As a consequence the ISM 
causes a stronger absorption signal, which more than compensates the increased 
contribution from the cold filament.

It should also be noted that finite resolution affects the metallicity of the 
filamentary gas, so that in the real Universe its metallicity may be higher than the values 
($\sim 10^{-4}-10^{-3} Z_{\odot}$) we report for the \mn\ simulation. Indeed, our simulation 
cannot resolve all the small galaxies that might pollute the pristine filamentary gas. 
Besides, it is well known that the energy from supernovae will
artificially be dissipated when simulations are run with an
insufficient level of resolution. Indeed, in such a case, supernovae only explode 
in dense grid cells, resulting in substantial radiative losses and negligible momentum 
transfer to the surrounding gas. Under these circumstances, metals cannot disperse properly.
For comparison, the ultra-high resolution \nut\ simulation indicates that the metallicity 
of the filamentary gas around a $\sim 10^{9}\msun $ galaxy can reach 
values up to $10^{-2} Z_{\odot}$ already at $z \simeq 7$. 
Yet, such an increase in metallicity would still be inconsequential for the
absorption spectra. Furthermore, we believe that the metallicity in the filaments is not likely to 
rise appreciably beyond these values, because most of the supernova ejecta escapes 
in a direction perpendicular to that of the elongated filament, 
which makes it difficult to efficiently enrich the filamentary gas with metals (Geen et al 2010, {\sl in prep.}). 
However, for the sake of completeness, we present in Fig.~\ref{fig:example} the case of a halo 
for which the gas outside $r > 0.1 r_{200}$ is arbitrarily assumed to have solar metallicity. We find that 
the absorption strength of the filament (red dotted line) is still
much smaller than the absorption produced by the 
galaxy's ISM.

\section{Conclusions and Discussion}

Cosmological simulations predict that high-z galaxies grow by  acquiring gas from 
cold streams \citep{dekel09}, but no observational confirmation has been obtained yet. 
Based on a statistical sample from the \mn\ simulation,
we argue that low-ionisation metal absorption features, such as  CII $\lambda1334$, 
arising from intervening cold filaments are extremely hard to distinguish from absorption 
by the ISM of high-z star-forming galaxies. This is primarily because 
the optical depth for the low-ionisation transition from cold filamentary gas is 
minuscule, compared with that of the ISM of the host galaxy. Moreover,
the filamentary absorption is not redshifted enough with respect to
the ISM absorption, so that the residual ISM absorption in the red
wing of the line is still prominent. This small optical depth of
filaments mainly finds its source in the intrinsically low densities and metallicities of the cold
gas when compared to those of the galaxy's ISM. Another factor is
the geometry of the flow which lowers the probability of detecting
filaments as their column density will rarely be maximised by being aligned with the line of sight. 

As an alternative to using a single galaxy, one
could probe circumgalactic regions by using a paired background galaxy. 
This method alleviates the importance of ISM absorption since when probing
circumgalactic regions in this way, the absorption by the ISM of the foreground 
galaxy will occur at a  different spatial position from that produced by filaments. 
Unfortunately, the rare occurrence of suitable foreground-background galaxy pairs  
makes it difficult to probe more than one line of sight per foreground
galaxy. As a result, high resolution individual spectra are hard to obtain and 
one has to resort to stacking the spectra of multiple galaxies \citep{steidel10}. 
Contrary to what might be expected, stacking will wash out the cold filament 
absorption signal since absorption by inflowing gas does not neatly separate 
from that caused by outflows as was the case when probing the circumgalactic 
region using single galaxies. Indeed, cold filament absorption against the 
background galaxy light will not only be redshifted, but also blueshifted as 
one expects that on average as many cold streams will be detected moving towards 
as away from the observer.  Absorption by outflows will also suffer the same fate.  
This will make it all the more difficult to argue whether the observed absorption 
features are driven by infalling gas or by outflows from high-z star-forming galaxies. 
Moreoever, the metal column density of cold filaments is minuscule, as already mentioned. 
The hydrogen column densities of the cold filaments are distributed around 
$10^{20} {\rm cm^{-2}}$ at these redshifts, yielding corresponding carbon column 
densities around $10^{16} {\rm cm^{-2}}$ if an average $Z=0.001 Z_{\odot}$ is used. 
Even if all carbon atoms are assumed to be eligible for the CII transition, 
the optical depth is only around $10^{-2}$ in the line. On the other hand, outflows are expected 
to be very metal rich, so that even though higher transitions like CIV are expected 
to dominate the absorption signal, the amount of CII absorption from these
outflows might still swamp that produced by the cold filaments.  

Finally, we have assessed possible numerical resolution issues on column density
and metallicity of the filaments using the very high resolution numerical
simulation \nut\ suite and found that our conclusions remain by and large unchanged. 

Based on these considerations, we conclude that the presence of the cold filament 
is difficult to disprove/prove with low-ionisation metal line absorption. 
Instead, the Lyman $\alpha$ emission route seems more promising to detect cold filaments, 
but the line profile is more sensitive to the kinematics of the intervening gas 
\citep{verhamme06,verhamme08}. Therefore this will require full blown radiative transfer 
calculations \citep[e.g.][]{faucher10b}.

\section*{Acknowledgements}
The \mn\ simulation was run at the Barcelona Supercomputing Center and 
the \nut\ simulations at the CINES and the STFC HPC Facility DiRAC (Oxford node). 
CP acknowledges support from a Leverhulme visiting professorship at the Astrophysics 
department of the University of Oxford, TK  from a Clarendon DPhil studentship. 
JD's research is supported by the Oxford Martin School, as is AS's along with Beecroft and STFC.
We also acknowledge support from the Franco-Korean PHC STAR program.

\end{document}